\documentclass[12pt,a4paper]{article}
\makeatletter
\def\eqnarray{%
\stepcounter{equation}%
\let\@currentlabel=\theequation
\global\@eqnswtrue
\global\@eqcnt\z@
\tabskip\@centering
\let\\=\@eqncr
$$\halign to \displaywidth\bgroup\@eqnsel\hskip\@centering
$\displaystyle\tabskip\z@{##}$&\global\@eqcnt\@ne
\hfil$\displaystyle{{}##{}}$\hfil
&\global\@eqcnt\tw@$\displaystyle\tabskip\z@{##}$\hfil
\tabskip\@centering&\llap{##}\tabskip\z@\cr}
\makeatother
\newcommand{\ket}[1]{{\vert{#1}\rangle}}
\newcommand{\bra}[1]{{\langle{#1}\vert}}

\newcommand{\fukuso}{{\mathbf C}}
\newcommand{\futon}{{\bf N}}
\newcommand{\seisu}{{\bf Z}}

\begin{document}

\title{\sl Cavity QED and Quantum Computation in the Strong Coupling Regime II
 : Two Atoms Case}
\author{
  Kazuyuki FUJII
  \thanks{E-mail address : fujii@yokohama-cu.ac.jp }\\
  Department of Mathematical Sciences\\
  Yokohama City University\\
  Yokohama, 236-0027\\
  Japan
  }
\date{}
\maketitle
%\thispagestyle{empty}
%
%
%  gaiyou
%
%
\begin{abstract}
  In this paper we study the behavior of (laser--cooled) m--atoms trapped in a 
  cavity interacting with a photon $\cdots$ Cavity QED $\cdots$ and attempt to 
  solve the Schr{\"o}dinger equation of this model in the strong coupling 
  regime. 
  In the case of m = 2 we construct Bell--Schr{\" o}dinger cat states (in our 
  terminology) and obtain with such bases some unitary transformations 
  by making use of the rotating wave approximation under new resonance 
  conditions containing the Bessel functions, which will be applied 
  to construct important quantum logic gates in Quantum Computation. 
  Moreover we propose in the case of m = 3 a crucial problem to solve on 
  Quantum Computation. 
\end{abstract}

%\newpage

%
%
%     Honbun
%
%
This paper is a continuation of \cite{KF7}, \cite{KF8}. 
We consider a unified model of the interaction of the two--level 
atoms and both the single radiation mode and external field (periodic usually)
in a cavity. We deal with the external field as a classical one. 
As a general introduction to this topic in Quantum Optics see \cite{AE}, 
\cite{MSIII}, \cite{C-HDG}. 
Our model is deeply related to the one of trapped ions in a cavity with 
the photon interaction (Cavity QED). 

In our model we are especially interested in the strong coupling regime, 
\cite{MFr2}, \cite{KF2}, \cite{KF3}. One of motivations is a recent very 
interesting experiment, \cite{NPT}. See \cite{C-HDG} and \cite{MFr3} 
as a general introduction. 

In \cite{MFr2} and \cite{KF2} we have treated the strong coupling regime of 
the interaction model of the two--level atom and the single radiation mode, 
and have given some explicit solutions under the resonance conditions and 
rotating wave approximations. 

On the other hand we want to add some external field (like Laser one) to 
the above model which will make the model more realistic (for example in 
Quantum Computation). Therefore we propose the unified model. 

We would like to solve our model in the strong coupling regime. 
Especially we want to show the existence of Rabi oscillations in this regime 
because the real purpose of a series of study (\cite{KF2}, \cite{KF3}, 
\cite{KF7}, \cite{KF8}) is an application to Quantum Computation (see 
\cite{KF1} as a brief introduction to it). 

Since in the previous paper \cite{KF8} we have treated one--atom case, 
so we deal with two--atoms case in this paper. 
To solve the Schr{\" o}dinger equation in this regime we construct (in our 
terminology) Bell--Schr{\" o}dinger cat states, and make use of rotating wave 
approximation under new resonance conditions containing the Bessel functions 
and etc, and obtain unitary transformations which are necessary to perform 
quantum logic gates. 

Our solutions might give a new insight into Quantum Optics or Condensed Matter 
Physics as well as Quantum Computation.

\vspace{10mm}
We here make some preparations for the latter. 
Let $\{\sigma_{1}, \sigma_{2}, \sigma_{3}\}$ be Pauli matrices and 
${\bf 1}_{2}$ a unit matrix : 
\begin{equation}
\sigma_{1} = 
\left(
  \begin{array}{cc}
    0& 1 \\
    1& 0
  \end{array}
\right), \quad 
\sigma_{2} = 
\left(
  \begin{array}{cc}
    0& -i \\
    i& 0
  \end{array}
\right), \quad 
\sigma_{3} = 
\left(
  \begin{array}{cc}
    1& 0 \\
    0& -1
  \end{array}
\right), \quad 
{\bf 1}_{2} = 
\left(
  \begin{array}{cc}
    1& 0 \\
    0& 1
  \end{array}
\right), 
\end{equation}
and 
$\sigma_{+} = (1/2)(\sigma_{1}+i\sigma_{2})$, 
$\sigma_{-} = (1/2)(\sigma_{1}-i\sigma_{2})$. 
Let $W$ be the Walsh--Hadamard matrix 
\begin{equation}
\label{eq:2-Walsh-Hadamard}
W=\frac{1}{\sqrt{2}}
\left(
  \begin{array}{cc}
    1& 1 \\
    1& -1
  \end{array}
\right)
=W^{-1}\ , 
\end{equation}
then we can diagonalize $\sigma_{1}$ as 
$
\sigma_{1}=W\sigma_{3}W^{-1}=W\sigma_{3}W
$
 by making use of this $W$. 
The eigenvalues of $\sigma_{1}$ is $\{1,-1\}$ with eigenvectors 
\begin{equation}
\label{eq:eigenvectors of sigma}
\ket{1}=\frac{1}{\sqrt{2}}
\left(
  \begin{array}{c}
    1 \\
    1
  \end{array}
\right), \quad 
\ket{-1}=\frac{1}{\sqrt{2}}
\left(
  \begin{array}{c}
    1 \\
    -1
  \end{array}
\right)
\quad \Longrightarrow \quad 
\ket{\lambda}=\frac{1}{\sqrt{2}}
\left(
  \begin{array}{c}
    1 \\
    \lambda
  \end{array}
\right).
\end{equation}

We have treated one atom with two--level, so we would like to generalize 
the method developped in \cite{KF2}, \cite{KF7} to $m$ atoms with two--level 
interacting both the single radiation mode and (classical) external fields 
like ($m$ atoms trapped in a cavity) 
\vspace{5mm} 
%%%%%%%%%%%%%%%%%%%%%%%%%%%%%%%%%%%%%%%%%%%%%%%%%%%%%%%%%%%%%%%
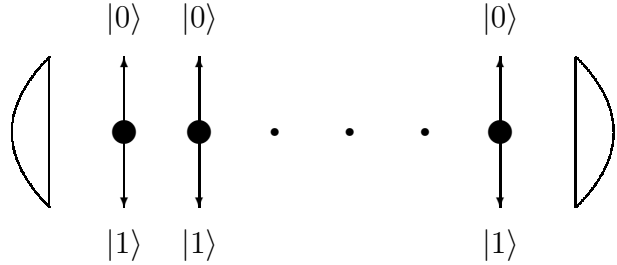
\begin{center}
\setlength{\unitlength}{1mm} 
\begin{picture}(120,40)(0,-10)
\bezier{200}(20,0)(10,10)(20,20)
\put(20,0){\line(0,1){20}}
\put(20,20){\makebox(20,10)[c]{$|0\rangle$}}
\put(30,10){\vector(0,1){10}}
\put(30,10){\vector(0,-1){10}}
\put(20,-10){\makebox(20,10)[c]{$|1\rangle$}}
\put(30,10){\circle*{3}}
\put(30,20){\makebox(20,10)[c]{$|0\rangle$}}
\put(40,10){\vector(0,1){10}}
\put(40,10){\vector(0,-1){10}}
\put(30,-10){\makebox(20,10)[c]{$|1\rangle$}}
\put(40,10){\circle*{3}}
\put(50,10){\circle*{1}}
\put(60,10){\circle*{1}}
\put(70,10){\circle*{1}}
\put(70,20){\makebox(20,10)[c]{$|0\rangle$}}
\put(80,10){\vector(0,1){10}}
\put(80,10){\vector(0,-1){10}}
\put(80,10){\circle*{3}}
\put(70,-10){\makebox(20,10)[c]{$|1\rangle$}}
\bezier{200}(90,0)(100,10)(90,20)
\put(90,0){\line(0,1){20}}
\end{picture}
\end{center}
%%%%%%%%%%%%%%%%%%%%%%%%%%%%%%%%%%%%%%%%%%%%%%%%%%%%%%%%%%%%%%%%
%
Then the unified Hamiltonian can be given as 
\begin{equation}
\label{eq:general-hamiltonian}
H_{L}
=\omega {\bf 1}_{M}\otimes L_{3} + 
g_{1}\sum_{j=1}^{m}\sigma_{1}^{(j)}\otimes (L_{+}+L_{-}) + 
\frac{\Delta}{2}\sum_{j=1}^{m}\sigma_{3}^{(j)}\otimes {\bf 1}_{L} + 
g_{2}\sum_{j=1}^{m}\mbox{cos}(\omega_{j}t+\phi_{j})\sigma_{1}^{(j)}\otimes 
{\bf 1}_{L}, 
\end{equation}
where $M=2^{m}$ and $\sigma_{k}^{(j)}$ ($k=1,\ 3$) is 
\[
\sigma_{k}^{(j)}=1_{2}\otimes \cdots \otimes 1_{2}\otimes \sigma_{k}\otimes 
1_{2}\otimes \cdots \otimes 1_{2}\ (j-\mbox{position}), 
\]
see \cite{KF7}. 
Here we are treating the following three cases at the same time : 
The notation $\{L_{+},L_{-},L_{3}\}$ means respectively 
\begin{equation}
\{L_{+},L_{-},L_{3}\}=
\left\{
\begin{array}{ll}
\mbox{(N)}\qquad \{a^{\dagger},a,N\}, \\
\mbox{(K)}\qquad \{K_{+},K_{-},K_{3}\}, \\
\mbox{(J)}\qquad \ \{J_{+},J_{-},J_{3}\}, 
\end{array}
\right.
\end{equation}
and $\{a^{\dagger},a,N\}$ is the generators of Heisenberg algebra, 
$\{K_{+},K_{-},K_{3}\}$ and $\{J_{+},J_{-},J_{3}\}$ are a set of 
generators of unitary representations of Lie algebras $su(1,1)$ and $su(2)$. 
They are usually constructed by making use of two harmonic oscillators 
(two--photons) $\{a_{1}, a_{1}^{\dagger}\},\ \{a_{2}, a_{2}^{\dagger}\}$ as 
\begin{eqnarray}
  \label{eq:schwinger-boson}
  su(1,1) &:&\quad
     K_+ = {a_1}^{\dagger}{a_2}^{\dagger},\ K_- = a_2 a_1,\ 
     K_3 = {1\over2}\left({a_1}^{\dagger}a_1 + {a_2}^{\dagger}a_2  + 1\right),
  \nonumber \\
  su(2) &:&\quad
     J_+ = {a_1}^{\dagger}a_2,\ J_- = {a_2}^{\dagger}a_1,\ 
     J_3 = {1\over2}\left({a_1}^{\dagger}a_1 - {a_2}^{\dagger}a_2\right).
  \nonumber
\end{eqnarray}
See for example \cite{KF2}, \cite{KF4} in detail.

We are interested in the strong coupling regime ($g_{1} \gg \Delta$), 
so we solve the Hamiltonian as follows (see \cite{KF7}). 
Let us transform (\ref{eq:general-hamiltonian}) into 
\begin{eqnarray}
\label{eq:perturbative-hamiltonian}
H_{L}
&=&\omega {\bf 1}_{M}\otimes L_{3} + 
\sum_{j=1}^{m}\sigma_{1}^{(j)}\otimes 
\left\{
g_{1}(L_{+}+L_{-}) + g_{2}\mbox{cos}(\omega_{j}t+\phi_{j}){\bf 1}_{L} 
\right\} + 
\frac{\Delta}{2}\sum_{j=1}^{m}\sigma_{3}^{(j)}\otimes {\bf 1}_{L} 
\nonumber \\ 
&\equiv& H_{0}+\frac{\Delta}{2}\sum_{j=1}^{m}\sigma_{3}^{(j)}\otimes 
{\bf 1}_{L},
\end{eqnarray}
where we have written $H_{0}$ instead of $H_{L0}$ for simplicity. 
First we diagonalize $H_{0}$. For that we set 
\[
{\bf W}=W\otimes W\otimes \cdots \otimes W\quad \in \quad U(M)
\]
for $W$ in (\ref{eq:2-Walsh-Hadamard}). Then it is not difficult to see 
(see \cite{KF2}, \cite{KF7}) 
\begin{eqnarray}
H_{0}
&=&({\bf W}\otimes {\bf 1}_{L})
\left[
{\bf 1}_{M}\otimes \omega L_{3} + 
\sum_{j=1}^{m}\sigma_{3}^{(j)}\otimes 
\left\{g_{1}(L_{+}+L_{-}) + g_{2} 
\mbox{cos}(\omega_{j}t+\phi_{j}){\bf 1}_{L}\right\}
\right]
({\bf W}^{-1}\otimes {\bf 1}_{L})   \nonumber \\
&=&({\bf W}\otimes {\bf 1}_{L})
\left[
{\bf 1}_{M}\otimes \omega L_{3} + 
g_{1}\sum_{j=1}^{m}\sigma_{3}^{(j)}\otimes(L_{+}+L_{-}) + 
g_{2}\sum_{j=1}^{m}\sigma_{3}^{(j)}\otimes 
\mbox{cos}(\omega_{j}t+\phi_{j}){\bf 1}_{L}
\right]
({\bf W}^{-1}\otimes {\bf 1}_{L})   \nonumber \\
&=&\sum_{\lambda}
\left(
\ket{\lambda}\otimes \mbox{e}^{-\frac{\Lambda x}{2}(L_{+}-L_{-})}
\right)
\left\{\Omega L_{3}+ g_{2}\sum_{j=1}^{m}\lambda_{j} 
\mbox{cos}(\omega_{j}t+\phi_{j}){\bf 1}_{L}\right\}
\left(
\bra{\lambda}\otimes \mbox{e}^{\frac{\Lambda x}{2}(L_{+}-L_{-})}
\right),   
\end{eqnarray}
where we have used the following 

\noindent{\bfseries Key Formulas}
\begin{eqnarray}
\label{eq:key-formula}
&&(N)\quad 
\omega a^{\dagger}a + g_{1}\Lambda (a^{\dagger}+a)
=\Omega\ \mbox{e}^{-\frac{\Lambda x}{2}(a^{\dagger}-a)}
\left(N-\frac{(g_{1}\Lambda)^2}{\omega^2}\right)
\mbox{e}^{\frac{\Lambda x}{2}(a^{\dagger}-a)} \nonumber \\
&&\quad \qquad \quad  \mbox{where}\quad 
\Omega=\omega, \quad  x=2g_{1}/\omega, \\
\label{eq:K-formula}
&&(K)\quad 
\omega K_{3} + g_{1}\Lambda (K_{+}+K_{-})
=\Omega\ \mbox{e}^{-\frac{\Lambda x}{2}(K_{+}-K_{-})} K_{3}\ 
\mbox{e}^{\frac{\Lambda x}{2}(K_{+}-K_{-})} \nonumber \\
&&\quad \qquad \quad  \mbox{where}\quad 
\Omega=\omega \sqrt{1-(2g_{1}\Lambda/\omega)^{2}}, \quad 
x=(1/\Lambda)\mbox{tanh}^{-1}(2g_{1}\Lambda/\omega), \\
\label{eq:J-formula}
&&(J)\quad
\omega J_{3} + g_{1}\Lambda (J_{+}+J_{-})
=\Omega\ \mbox{e}^{-\frac{\Lambda x}{2}(J_{+}-J_{-})} J_{3}\ 
\mbox{e}^{\frac{\Lambda x}{2}(J_{+}-J_{-})} \nonumber \\
&&\quad \qquad \quad  \mbox{where}\quad 
\Omega=\omega \sqrt{1+(2g_{1}\Lambda/\omega)^{2}}, \quad 
x=(1/\Lambda)\mbox{tan}^{-1}(2g_{1}\Lambda/\omega), 
\end{eqnarray}
and have used concise notations 
$
\lambda=(\lambda_{1},\cdots,\lambda_{j-1},\lambda_{j},\lambda_{j+1},
\cdots,\lambda_{m})
$ 
and 
\[
\sum_{\lambda}=
\sum_{\lambda_{1}=\pm 1}\sum_{\lambda_{2}=\pm 1}\cdots 
\sum_{\lambda_{m}=\pm 1}\quad ;\quad 
\ket{\lambda}=\ket{\lambda_{1}}\otimes \ket{\lambda_{2}}\otimes \cdots 
\otimes \ket{\lambda_{m}}\quad ;\quad 
\Lambda\equiv \Lambda(\lambda)=\sum_{j=1}^{m}\lambda_{j} 
\]
for $\lambda_{j}=\pm 1$. We leave the proof of Key Formulas to the 
readers. 

That is, we could diagonalize the Hamiltonian $H_{0}$. 
Its eigenvalues $\{E_{n}(t)\}$ and eigenvectors $\{\ket{\{\lambda, n\}}\}$ 
are given respectively 
\begin{eqnarray}
\label{eq:Eigenvalues-Eigenvectors}
&&(E_{n}(t),\ \ket{\{\lambda, n\}})= \nonumber \\
&&\left\{
\begin{array}{ll}
(N)\quad \Omega (-\frac{g_{1}^2}{\omega^2}\Lambda^{2}+n)+
g_{2}\sum_{j=1}^{m}\lambda_{j} \mbox{cos}(\omega_{j}t+\phi_{j})
, \quad \ket{\lambda}\otimes 
\mbox{e}^{-\frac{\Lambda x}{2}(a^{\dagger}-a)}\ket{n}, \\
(K)\quad \Omega (K+n)+
g_{2}\sum_{j=1}^{m}\lambda_{j} \mbox{cos}(\omega_{j}t+\phi_{j})
,\quad \quad 
\ket{\lambda}\otimes 
\mbox{e}^{-\frac{\Lambda x}{2}(K_{+}-K_{-})}\ket{K,n}, \\
(J)\quad \ \Omega (-J+n)+
g_{2}\sum_{j=1}^{m}\lambda_{j} \mbox{cos}(\omega_{j}t+\phi_{j})
,\quad \ 
\ket{\lambda}\otimes 
\mbox{e}^{-\frac{\Lambda x}{2}(J_{+}-J_{-})}\ket{J,n} \\
\end{array}
\right.
\end{eqnarray}
for $n \in \futon \cup \{0\}$, where 
$
E_{n}(t)\equiv 
E_{n}(\lambda)+g_{2}\sum_{j=1}^{m}\lambda_{j} \mbox{cos}(\omega_{j}t+\phi_{j})
$. 
Then $H_{0}$ above can be written as 
\begin{eqnarray}
H_{0}
&=&\sum_{\lambda}\sum_{n}E_{n}(t)\ket{\{\lambda, n\}}\bra{\{\lambda, n\}}
\nonumber \\
&=&\sum_{\lambda}\sum_{n}
\left\{
E_{n}(\lambda)+g_{2}\sum_{j=1}^{m}\lambda_{j} \mbox{cos}(\omega_{j}t+\phi_{j})
\right\}\ket{\{\lambda, n\}}\bra{\{\lambda, n\}}.  \nonumber 
\end{eqnarray}

Up to this stage we could treat L = (N), (K), (J) at the same time. However 
in the following it is very difficult to treat them at the same time, so 
we shall focus our attention to the simplest case L = (N)
\footnote{The main difficulty for L = (K) and (J) is the $\lambda$ dependence 
of $x$, see (\ref{eq:K-formula}) and (\ref{eq:J-formula}).}. 
We leave the remaining cases as a future (challenging) task. 

\par \vspace{3mm} 
Next we would like to solve the following Schr{\"o}dinger equation : 
\begin{equation}
\label{eq:full-equation}
i\frac{d}{dt}\Psi=H_{L}\Psi=
\left(
H_{0}+
\frac{\Delta}{2}\sum_{j=1}^{m}\sigma_{3}^{(j)}\otimes {\bf 1}_{L}
\right)\Psi. 
\end{equation}
To solve this equation we appeal to the method of constant variation. 
First let us solve 
$
i\frac{d}{dt}\Psi=H_{0}\Psi, 
$
which general solution is given by 
$
\label{eq:partial-solution}
  \Psi(t)=U_{0}(t)\Psi_{0}
$, 
where $\Psi_{0}$ is a constant state and 
\begin{equation}
\label{eq:Basic-Unitary}
U_{0}(t)=\sum_{\lambda}\sum_{n}
\mbox{e}^{-i\left\{tE_{n}(\lambda)+\sum_{j=1}^{m}\lambda_{j}(g_{2}/\omega_{j})
sin(\omega_{j}t+\phi_{j})\right\}}
\ket{\{\lambda, n\}}\bra{\{\lambda, n\}}.
\end{equation}
The method of constant variation goes as follows. Changing like 
$
\Psi_{0} \longrightarrow \Psi_{0}(t),
$ 
we have 
\begin{equation}
\label{eq:sub-equation}
i\frac{d}{dt}\Psi_{0}
=\frac{\Delta}{2}{U_0}^{\dagger}
\left(\sum_{j=1}^{m}\sigma_{3}^{(j)}\otimes {\bf 1}_{L}\right)
{U_0}\Psi_{0} 
\equiv \frac{\Delta}{2}H_{F}\Psi_{0}. 
\end{equation}
Then making use of the fact 
$\bra{\lambda_{j}}\sigma_{3}=\bra{-\lambda_{j}}$ and after long calculations, 
$H_{F}$ becomes 
\begin{eqnarray}
H_{F}=      
\sum_{\lambda}\sum_{n,n'}\sum_{j=1}^{m}
&&\mbox{e}^{
i\left\{t\omega (n-n')+t\omega x^{2}(1-\lambda_{j}\Lambda(\lambda))
+2\lambda_{j}(g_{2}/\omega_{j})sin(\omega_{j}t+\phi_{j})
\right\}}\times     \nonumber \\
&&\bra{n}\mbox{e}^{{\lambda_{j} x}(a^{\dagger}-a)}\ket{n'}
\ket{\{\lambda, n\}}\bra{ \{\lambda, n'\}_{(j)} }, 
\end{eqnarray}
where we have used the compact notations 
$
\lambda_{(j)}=(\lambda_{1},\cdots,\lambda_{j-1},-\lambda_{j},\lambda_{j+1},
\cdots,\lambda_{m})
$ 
and 
\begin{eqnarray}
\bra{\lambda_{(j)}}&=&
\bra{\lambda_{1}}\otimes \cdots \otimes \bra{-\lambda_{j}}\otimes \cdots 
\otimes \bra{\lambda_{m}}, \nonumber \\
\bra{\{\lambda, n\}_{(j)}}&=&\bra{\lambda_{(j)}}\otimes 
\bra{n}
\mbox{exp}\left\{\frac{\Lambda(\lambda_{(j)}) x}{2}(a^{\dagger}-a)\right\}, 
\nonumber 
\end{eqnarray}
and 
\[
\Lambda(\lambda_{(j)}) = \lambda_{1}+\cdots +\lambda_{j-1}-\lambda_{j}+
\lambda_{j+1}+\cdots +\lambda_{m}
=\Lambda(\lambda)-2\lambda_{j} 
\]
and 
\begin{eqnarray}
E_{n}(\lambda)-E_{n'}(\lambda_{(j)})
&=&
\omega\left\{-\frac{x^2}{4}\Lambda(\lambda)^{2}+n \right\}-
\omega\left\{-\frac{x^2}{4}\Lambda(\lambda_{(j)})^{2}+n' \right\}
\nonumber \\
&=&
\omega(n-n')+\omega x^{2}(1-\lambda_{j}\Lambda(\lambda)). \nonumber 
\end{eqnarray}
In the following we set for simplicity $\phi_{j}=0$ and 
\begin{equation}
\label{eq:time-depend-F}
\Theta_{j}(t)\equiv g_{2}\frac{\mbox{sin}(\omega_{j}t)}{\omega_{j}}
\quad \mbox{for}\quad 1 \leq j \leq m, 
\end{equation}
then 
\begin{equation}
H_{F}=      
\sum_{\lambda}\sum_{n,n'}\sum_{j=1}^{m}
\mbox{e}^{i\left\{t\omega (n-n')+t\omega x^{2}(1-\lambda_{j}\Lambda(\lambda))
+2\lambda_{j}\Theta_{j}(t)\right\}} 
\bra{n}\mbox{e}^{{\lambda_{j} x}(a^{\dagger}-a)}\ket{n'}
\ket{\{\lambda, n\}}\bra{ \{\lambda, n'\}_{(j)} }.
\end{equation}

Here we divide $H_{F}$ into two parts\ 
$
H_{F}={H_{F}}^{'}+{H_{F}}^{''}
$
\ where 
\begin{equation}
\label{eq:Second-Hamiltonian-1}
{H_{F}}^{'}=      
\sum_{\lambda}\sum_{n}\sum_{j=1}^{m}
\mbox{e}^{i\left\{t\omega x^{2}(1-\lambda_{j}\Lambda(\lambda))
+2\lambda_{j}\Theta_{j}(t)\right\}} 
\bra{n}\mbox{e}^{{\lambda_{j} x}(a^{\dagger}-a)}\ket{n}
\ket{\{\lambda, n\}}\bra{ \{\lambda, n\}_{(j)} },
\end{equation}
and 
\begin{equation}
\label{eq:Second-Hamiltonian-2}
{H_{F}}^{''}=      
\sum_{\lambda}\sum_{\stackrel{\scriptstyle n,n'}{n\ne n'}}
\sum_{j=1}^{m}
\mbox{e}^{i\left\{t\omega (n-n')+t\omega x^{2}(1-\lambda_{j}\Lambda(\lambda))
+2\lambda_{j}\Theta_{j}(t)\right\}} 
\bra{n}\mbox{e}^{{\lambda_{j} x}(a^{\dagger}-a)}\ket{n'}
\ket{\{\lambda, n\}}\bra{ \{\lambda, n'\}_{(j)} }.
\end{equation}

Noting 
$
\bra{n}\mbox{e}^{x(a^{\dagger}-a)}\ket{n}=
\bra{n}\mbox{e}^{-x(a^{\dagger}-a)}\ket{n}
$
($\lambda_{j}=\pm 1$) 
by the results in section 3 of \cite{KF2}, ${H_{F}}^{'}$ can be 
written as
\begin{equation}
\label{eq:Second-Hamiltonian-1-modify}
{H_{F}}^{'}=\sum_{n}\bra{n}\mbox{e}^{x(a^{\dagger}-a)}\ket{n} 
\left\{
\sum_{\lambda}\sum_{j=1}^{m}
\mbox{e}^{ i\left\{t\omega x^{2}(1-\lambda_{j}\Lambda(\lambda))
+2\lambda_{j}\Theta_{j}(t)\right\} } 
\ket{\{\lambda, n\}}\bra{\{\lambda, n\}_{(j)}}
\right\}. 
\end{equation}
For us it is not easy to expand the equation above, so 
we consider the special case. 
\begin{flushleft}
{\sl {\Large Case of m=2}}\ {\Large :}
\end{flushleft}
\begin{eqnarray}
\label{eq:expansion}
&&\left\{\sum_{\lambda}\sum_{j=1}^{2} \cdots \right\}  \nonumber \\
=\ 
&&
\mbox{e}^{i\{-t\omega x^{2}+2\Theta_{1}(t)\}}
\ket{\{1,1, n\}}\bra{\{1,1, n\}_{(1)}}+
\mbox{e}^{i\{-t\omega x^{2}+2\Theta_{2}(t)\}}
\ket{\{1,1, n\}}\bra{\{1,1, n\}_{(2)}}+      \nonumber \\
&&
\mbox{e}^{i\{t\omega x^{2}+2\Theta_{1}(t)\}}
\ket{\{1,-1, n\}}\bra{\{1,-1, n\}_{(1)}}+    
\mbox{e}^{i\{t\omega x^{2}-2\Theta_{2}(t)\}}
\ket{\{1,-1, n\}}\bra{\{1,-1, n\}_{(2)}}+    \nonumber \\
&&
\mbox{e}^{i\{t\omega x^{2}-2\Theta_{1}(t)\}}
\ket{\{-1,1, n\}}\bra{\{-1,1, n\}_{(1)}}+
\mbox{e}^{i\{t\omega x^{2}+2\Theta_{2}(t)\}}
\ket{\{-1,1, n\}}\bra{\{-1,1, n\}_{(2)}}+    \nonumber \\
&&
\mbox{e}^{i\{-t\omega x^{2}-2\Theta_{1}(t)\}}
\ket{\{-1,-1, n\}}\bra{\{-1,-1, n\}_{(1)}}+
\mbox{e}^{i\{-t\omega x^{2}-2\Theta_{2}(t)\}}
\ket{\{-1,-1, n\}}\bra{\{-1,-1, n\}_{(2)}}  \nonumber \\
=\ 
&&
\mbox{e}^{i\{-t\omega x^{2}+2\Theta_{1}(t)\}}
\ket{\{1,1, n\}}\bra{\{1,1, n\}_{(1)}}+
\mbox{e}^{i\{-t\omega x^{2}+2\Theta_{2}(t)\}}
\ket{\{1,1, n\}}\bra{\{1,1, n\}_{(2)}}+      \nonumber \\
&&
\mbox{e}^{i\{-t\omega x^{2}-2\Theta_{2}(t)\}}
\ket{\{-1,-1, n\}}\bra{\{-1,-1, n\}_{(2)}}+  
\mbox{e}^{i\{-t\omega x^{2}-2\Theta_{1}(t)\}}
\ket{\{-1,-1, n\}}\bra{\{-1,-1, n\}_{(1)}}+  \nonumber \\
&&
\mbox{e}^{i\{t\omega x^{2}-2\Theta_{2}(t)\}}
\ket{\{1,-1, n\}}\bra{\{1,-1, n\}_{(2)}}+    
\mbox{e}^{i\{t\omega x^{2}+2\Theta_{1}(t)\}}
\ket{\{1,-1, n\}}\bra{\{1,-1, n\}_{(1)}}+    \nonumber \\
&&
\mbox{e}^{i\{t\omega x^{2}-2\Theta_{1}(t)\}}
\ket{\{-1,1, n\}}\bra{\{-1,1, n\}_{(1)}}+
\mbox{e}^{i\{t\omega x^{2}+2\Theta_{2}(t)\}}
\ket{\{-1,1, n\}}\bra{\{-1,1, n\}_{(2)}}    \nonumber \\
=\ 
&&
\mbox{e}^{i\{-t\omega x^{2}+2\Theta_{1}(t)\}}
\ket{\{1,1, n\}}\bra{\{-1,1, n\}}+
\mbox{e}^{i\{-t\omega x^{2}+2\Theta_{2}(t)\}}
\ket{\{1,1, n\}}\bra{\{1,-1, n\}}+      \nonumber \\
&&
\mbox{e}^{i\{-t\omega x^{2}-2\Theta_{2}(t)\}}
\ket{\{-1,-1, n\}}\bra{\{-1,1, n\}}+  
\mbox{e}^{i\{-t\omega x^{2}-2\Theta_{1}(t)\}}
\ket{\{-1,-1, n\}}\bra{\{1,-1, n\}}+  \nonumber \\
&&
\mbox{e}^{i\{t\omega x^{2}-2\Theta_{1}(t)\}}
\ket{\{-1,1, n\}}\bra{\{1,1, n\}}+
\mbox{e}^{i\{t\omega x^{2}-2\Theta_{2}(t)\}}
\ket{\{1,-1, n\}}\bra{\{1,1, n\}}+    \nonumber \\
&&
\mbox{e}^{i\{t\omega x^{2}+2\Theta_{2}(t)\}}
\ket{\{-1,1, n\}}\bra{\{-1,-1, n\}}+  
\mbox{e}^{i\{t\omega x^{2}+2\Theta_{1}(t)\}}
\ket{\{1,-1, n\}}\bra{\{-1,-1, n\}},   
\end{eqnarray}
where let us once more remind 
\begin{eqnarray}
\label{eq:usual-basis}
\ket{\{1,1, n\}}&=&
\ket{1}\otimes \ket{1}\otimes \mbox{e}^{-x(a^{\dagger}-a)}\ket{n}, \nonumber \\
\ket{\{1,-1, n\}}&=&
\ket{1}\otimes \ket{-1}\otimes \ket{n}, \nonumber \\
\ket{\{-1,1, n\}}&=&
\ket{-1}\otimes \ket{1}\otimes \ket{n}, \nonumber \\
\ket{\{-1,-1, n\}}&=&
\ket{-1}\otimes \ket{-1}\otimes \mbox{e}^{x(a^{\dagger}-a)}\ket{n}. 
\end{eqnarray}

Here we want to solve the equation 
$i(d/dt)\Psi_{0}={H_{F}}^{'}\Psi_{0}$ completely, 
however it is very hard (see for example \cite{MFr1}, \cite{MFr4}, 
\cite{BaWr}, \cite{SGD}). Therefore let us appeal to a perturbation theory. 
Now we use the well--known formula 
\begin{equation}
\label{eq:Bessels}
\mbox{e}^{2i\lambda_{j}\Theta_{j}(t)}=\sum_{\alpha_{j}\in \seisu}
J_{\alpha_{j}}(2\lambda_{j} g_{2}/\omega_{j})
\mbox{e}^{i \alpha_{j}\omega_{j}t}
=J_{0}(2g_{2}/\omega_{j})+\sum_{\alpha_{j}\ne 0}
J_{\alpha_{j}}(2\lambda_{j} g_{2}/\omega_{j})
\mbox{e}^{i \alpha_{j}\omega_{j} t},
\end{equation}
where $J_{\alpha}(x)$ are the Bessel functions. 
For a further simplicity we set $2g_{2}/\omega_{j}=\Gamma_{j}$.

Here we define a kind of extended Bell states from (\ref{eq:usual-basis}) 
($\ket{1}\longleftrightarrow \ket{0},\ \ket{-1}\longleftrightarrow \ket{1}$ 
in a usual manner) 
\begin{eqnarray}
\ket{\{\Phi_{1},n\}}&=&
\frac{1}{\sqrt{2}}\left(\ket{\{1,1, n\}}+\ket{\{-1,-1, n\}}\right), 
\nonumber \\
\ket{\{\Phi_{2},n\}}&=&
\frac{1}{\sqrt{2}}\left(\ket{\{1,1, n\}}-\ket{\{-1,-1, n\}}\right), 
\nonumber \\
\ket{\{\Phi_{3},n\}}&=&
\frac{1}{\sqrt{2}}\left(\ket{\{-1,1, n\}}+\ket{\{1,-1, n\}}\right), 
\nonumber \\
\ket{\{\Phi_{4},n\}}&=&
\frac{1}{\sqrt{2}}\left(\ket{\{-1,1, n\}}-\ket{\{1,-1, n\}}\right).
\end{eqnarray}

\par \noindent
A comment is in order. We want to call these states (bases) 
Bell--Schr{\" o}dinger cat states. 

\par \noindent
Conversely we have 
\begin{eqnarray}
\ket{\{1,1, n\}}&=&
\frac{1}{\sqrt{2}}\left(\ket{\{\Phi_{1},n\}}+\ket{\{\Phi_{2},n\}}\right), 
\nonumber \\
\ket{\{-1,-1, n\}}&=&
\frac{1}{\sqrt{2}}\left(\ket{\{\Phi_{1},n\}}-\ket{\{\Phi_{2},n\}}\right), 
\nonumber \\
\ket{\{-1,1, n\}}&=&
\frac{1}{\sqrt{2}}\left(\ket{\{\Phi_{3},n\}}+\ket{\{\Phi_{4},n\}}\right), 
\nonumber \\
\ket{\{1,-1, n\}}&=&
\frac{1}{\sqrt{2}}\left(\ket{\{\Phi_{3},n\}}-\ket{\{\Phi_{4},n\}}\right). 
\end{eqnarray}

From (\ref{eq:sub-equation}) 
\[
i\frac{d}{dt}\Psi_{0}
=\frac{\Delta}{2}H_{F}\Psi_{0}
=\frac{\Delta}{2}\left(H_{F}^{'}+H_{F}^{''}\right)\Psi_{0}.
\]
However we are not interested in a transition from one state to 
an another one from the lesson obtaining one qubit case in \cite{KF8}, 
so we can remove $H_{F}^{''}$ ($n\neq n^{'}$) from the Hamiltonian. 
Next let us transform $H_{F}^{'}$. 
From (\ref{eq:expansion}) and (\ref{eq:Bessels}) if we set 
\[
\left\{\sum_{\lambda}\sum_{j=1}^{2} \cdots \right\}\equiv 
{K_{0F}}^{'}+{K_{1F}}^{'}\ : 
\]
then we obtain after some algebras 
\begin{eqnarray}
{K_{0F}}^{'}
=
&&\left(J_{0}(\Gamma_{1})+J_{0}(\Gamma_{2})\right)
\left\{
\mbox{e}^{-it\omega x^{2}}\ket{\{\Phi_{1},n\}}\bra{\{\Phi_{3},n\}}+
\mbox{e}^{it\omega x^{2}}\ket{\{\Phi_{3},n\}}\bra{\{\Phi_{1},n\}}
\right\}+   \nonumber \\
&&\left(J_{0}(\Gamma_{1})-J_{0}(\Gamma_{2})\right)
\left\{
\mbox{e}^{-it\omega x^{2}}\ket{\{\Phi_{2},n\}}\bra{\{\Phi_{4},n\}}+
\mbox{e}^{it\omega x^{2}}\ket{\{\Phi_{4},n\}}\bra{\{\Phi_{2},n\}}
\right\},
\end{eqnarray}
and
\begin{eqnarray}
&&{K_{1F}}^{'}=  \nonumber \\
&&\mbox{e}^{-it\omega x^{2}}\sum_{\alpha\neq 0}
\left\{
\frac{J_{\alpha}(\Gamma_{1})+J_{\alpha}(-\Gamma_{1})}{2}
\mbox{e}^{it\alpha\omega_{1}}+
\frac{J_{\alpha}(\Gamma_{2})+J_{\alpha}(-\Gamma_{2})}{2}
\mbox{e}^{it\alpha\omega_{2}}
\right\}
\ket{\{\Phi_{1},n\}}\bra{\{\Phi_{3},n\}}+    \nonumber \\
&&\mbox{e}^{-it\omega x^{2}}\sum_{\alpha\neq 0}
\left\{
\frac{J_{\alpha}(\Gamma_{1})-J_{\alpha}(-\Gamma_{1})}{2}
\mbox{e}^{it\alpha\omega_{1}}-
\frac{J_{\alpha}(\Gamma_{2})-J_{\alpha}(-\Gamma_{2})}{2}
\mbox{e}^{it\alpha\omega_{2}}
\right\}
\ket{\{\Phi_{1},n\}}\bra{\{\Phi_{4},n\}}+    \nonumber \\
&&\mbox{e}^{-it\omega x^{2}}\sum_{\alpha\neq 0}
\left\{
\frac{J_{\alpha}(\Gamma_{1})-J_{\alpha}(-\Gamma_{1})}{2}
\mbox{e}^{it\alpha\omega_{1}}+
\frac{J_{\alpha}(\Gamma_{2})-J_{\alpha}(-\Gamma_{2})}{2}
\mbox{e}^{it\alpha\omega_{2}}
\right\}
\ket{\{\Phi_{2},n\}}\bra{\{\Phi_{3},n\}}+    \nonumber \\
&&\mbox{e}^{-it\omega x^{2}}\sum_{\alpha\neq 0}
\left\{
\frac{J_{\alpha}(\Gamma_{1})+J_{\alpha}(-\Gamma_{1})}{2}
\mbox{e}^{it\alpha\omega_{1}}-
\frac{J_{\alpha}(\Gamma_{2})+J_{\alpha}(-\Gamma_{2})}{2}
\mbox{e}^{it\alpha\omega_{2}}
\right\}
\ket{\{\Phi_{2},n\}}\bra{\{\Phi_{4},n\}}+    \nonumber \\
&&\mbox{e}^{it\omega x^{2}}\sum_{\alpha\neq 0}
\left\{
\frac{J_{\alpha}(-\Gamma_{1})+J_{\alpha}(\Gamma_{1})}{2}
\mbox{e}^{it\alpha\omega_{1}}+
\frac{J_{\alpha}(-\Gamma_{2})+J_{\alpha}(\Gamma_{2})}{2}
\mbox{e}^{it\alpha\omega_{2}}
\right\}
\ket{\{\Phi_{3},n\}}\bra{\{\Phi_{1},n\}}+    \nonumber \\
&&\mbox{e}^{it\omega x^{2}}\sum_{\alpha\neq 0}
\left\{
\frac{J_{\alpha}(-\Gamma_{1})-J_{\alpha}(\Gamma_{1})}{2}
\mbox{e}^{it\alpha\omega_{1}}-
\frac{J_{\alpha}(-\Gamma_{2})-J_{\alpha}(\Gamma_{2})}{2}
\mbox{e}^{it\alpha\omega_{2}}
\right\}
\ket{\{\Phi_{4},n\}}\bra{\{\Phi_{1},n\}}+    \nonumber \\
&&\mbox{e}^{it\omega x^{2}}\sum_{\alpha\neq 0}
\left\{
\frac{J_{\alpha}(-\Gamma_{1})-J_{\alpha}(\Gamma_{1})}{2}
\mbox{e}^{it\alpha\omega_{1}}+
\frac{J_{\alpha}(-\Gamma_{2})-J_{\alpha}(\Gamma_{2})}{2}
\mbox{e}^{it\alpha\omega_{2}}
\right\}
\ket{\{\Phi_{3},n\}}\bra{\{\Phi_{2},n\}}+    \nonumber \\
&&\mbox{e}^{it\omega x^{2}}\sum_{\alpha\neq 0}
\left\{
\frac{J_{\alpha}(-\Gamma_{1})+J_{\alpha}(\Gamma_{1})}{2}
\mbox{e}^{it\alpha\omega_{1}}-
\frac{J_{\alpha}(-\Gamma_{2})+J_{\alpha}(\Gamma_{2})}{2}
\mbox{e}^{it\alpha\omega_{2}}
\right\}
\ket{\{\Phi_{4},n\}}\bra{\{\Phi_{2},n\}}, 
\end{eqnarray}
where we have used the basic relations 
\[
J_{\alpha}(-x)=(-1)^{\alpha}J_{\alpha}(x)=J_{-\alpha}(x).
\]
Therefore 
\begin{equation}
\label{eq:sub-sub-equation}
i\frac{d}{dt}\Psi_{0}
=\frac{\Delta}{2}H_{F}^{'}\Psi_{0}
=\sum_{n}
 \left\{
 \frac{\Delta}{2}\bra{n}\mbox{e}^{x(a^{\dagger}-a)}\ket{n}{K_{0F}}^{'}+
 \frac{\Delta}{2}\bra{n}\mbox{e}^{x(a^{\dagger}-a)}\ket{n}{K_{1F}}^{'}
 \right\}\Psi_{0}. 
\end{equation}
In the following we treat 
$\frac{\Delta}{2}\bra{n}\mbox{e}^{x(a^{\dagger}-a)}\ket{n}{K_{0F}}^{'}$ 
a unperturbed Hamiltonian and the remaining a perturbed one. 
If we set for simplicity 
\begin{eqnarray}
E_{\Delta,n,+}
&=&\frac{\Delta}{2}\bra{n}\mbox{e}^{x(a^{\dagger}-a)}\ket{n}
\left(J_{0}(\Gamma_{1})+J_{0}(\Gamma_{2})\right),  \nonumber \\
E_{\Delta,n,-}
&=&\frac{\Delta}{2}\bra{n}\mbox{e}^{x(a^{\dagger}-a)}\ket{n}
\left(J_{0}(\Gamma_{1})-J_{0}(\Gamma_{2})\right),  
\end{eqnarray}
then it is easy to solve 
\begin{eqnarray}
i\frac{d}{dt}\Psi_{0}
=\sum_{n}&&\frac{\Delta}{2}
 \left\{
 \bra{n}\mbox{e}^{x(a^{\dagger}-a)}\ket{n}{K_{0F}}^{'}
 \right\}\Psi_{0}       \nonumber \\
=\sum_{n}
&&
\left[E_{\Delta,n,+}
\left\{
\mbox{e}^{-it\omega x^{2}}\ket{\{\Phi_{1},n\}}\bra{\{\Phi_{3},n\}}+
\mbox{e}^{it\omega x^{2}}\ket{\{\Phi_{3},n\}}\bra{\{\Phi_{1},n\}}
\right\}+  \right.   \nonumber \\
&&\ \left. E_{\Delta,n,-}
\left\{
\mbox{e}^{-it\omega x^{2}}\ket{\{\Phi_{2},n\}}\bra{\{\Phi_{4},n\}}+
\mbox{e}^{it\omega x^{2}}\ket{\{\Phi_{4},n\}}\bra{\{\Phi_{2},n\}}
\right\}
\right]\Psi_{0},     \nonumber 
\end{eqnarray}
see Appendix. 

\par \noindent 
As an application to Quantum Computation it is sufficient for us to 
consider one excited state. 
By making use of the method of constant variation again we can set 
$\Psi_{0}(t)$ as 
\begin{eqnarray}
\label{eq:full-ansatz}
\Psi_{0}(t)=
&&(u_{n,11}c_{n,1}(t)+u_{n,13}c_{n,3}(t))\ket{\{\Phi_{1},n\}}+
(u_{n,22}c_{n,2}(t)+u_{n,24}c_{n,4}(t))\ket{\{\Phi_{2},n\}}+
\nonumber \\
&&(u_{n,31}c_{n,1}(t)+u_{n,33}c_{n,3}(t))\ket{\{\Phi_{3},n\}}+
(u_{n,42}c_{n,2}(t)+u_{n,44}c_{n,4}(t))\ket{\{\Phi_{4},n\}}
\end{eqnarray}
for some fixed $n$, where 
\begin{eqnarray}
\label{eq:special-unitary-1}
U_{n}(t)&=&
\left(
  \begin{array}{cc}
    u_{n,11}& u_{n,13}\\
    u_{n,31}& u_{n,33}
  \end{array}
\right)=
\left(
  \begin{array}{cc}
    1& \\
     & \mbox{e}^{it\omega x^{2}}
  \end{array}
\right)
\mbox{exp}\left\{-it
\left(
  \begin{array}{cc}
    0& E_{\Delta,n,+} \\
    E_{\Delta,n,+}& \omega x^{2}
  \end{array}
\right)
\right\}, \\
\label{eq:special-unitary-2}
V_{n}(t)&=&
\left(
  \begin{array}{cc}
    u_{n,22}& u_{n,24}\\
    u_{n,42}& u_{n,44}
  \end{array}
\right)=
\left(
  \begin{array}{cc}
    1& \\
     & \mbox{e}^{it\omega x^{2}}
  \end{array}
\right)
\mbox{exp}\left\{-it
\left(
  \begin{array}{cc}
    0& E_{\Delta,n,-} \\
    E_{\Delta,n,-}& \omega x^{2}
  \end{array}
\right)
\right\}
\end{eqnarray}
from the appendix. 

Then substituting (\ref{eq:full-ansatz}) into (\ref{eq:sub-sub-equation}) 
and some tedious calculations lead us to 
\begin{eqnarray}
i\frac{d}{dt}
\left(
  \begin{array}{c}
    c_{n,1}\\
    c_{n,3}\\
    c_{n,2}\\
    c_{n,4}
  \end{array}
\right) 
=&&         
\left(
  \begin{array}{cccc}
    u_{n,11}& u_{n,13}&     &   \\
    u_{n,31}& u_{n,33}&     &   \\
            &         & u_{n,22}& u_{n,24} \\
            &         & u_{n,42}& u_{n,44} 
  \end{array}
\right)^{-1}
\left(
  \begin{array}{cccc}
           0&       A&        0&  B       \\
    {\bar A}&       0& {\bar C}&  0       \\
           0&       C&        0&  D       \\
    {\bar B}&       0& {\bar D}&  0
  \end{array}
\right)\times  \nonumber \\
&&\left(
  \begin{array}{cccc}
    u_{n,11}& u_{n,13}&   &  \\
    u_{n,31}& u_{n,33}&   &  \\
            &         & u_{n,22}& u_{n,24} \\
            &         & u_{n,42}& u_{n,44} \\
  \end{array}
\right)
\left(
  \begin{array}{c}
    c_{n,1}\\
    c_{n,3}\\
    c_{n,2}\\
    c_{n,4}
  \end{array}
\right)
\end{eqnarray}
with 
\begin{eqnarray}
\label{eq:coefficients}
A&=&\mbox{e}^{-it\omega x^{2}}
\frac{\Delta}{2}\bra{n}\mbox{e}^{x(a^{\dagger}-a)}\ket{n}\sum_{\alpha\neq 0}
\left\{
\frac{J_{\alpha}(\Gamma_{1})+J_{\alpha}(-\Gamma_{1})}{2}
\mbox{e}^{it\alpha\omega_{1}}+
\frac{J_{\alpha}(\Gamma_{2})+J_{\alpha}(-\Gamma_{2})}{2}
\mbox{e}^{it\alpha\omega_{2}}
\right\}  \nonumber \\
&\equiv& \mbox{e}^{-it\omega x^{2}}A_{0}, \nonumber \\
B&=&\mbox{e}^{-it\omega x^{2}}
\frac{\Delta}{2}\bra{n}\mbox{e}^{x(a^{\dagger}-a)}\ket{n}\sum_{\alpha\neq 0}
\left\{
\frac{J_{\alpha}(\Gamma_{1})-J_{\alpha}(-\Gamma_{1})}{2}
\mbox{e}^{it\alpha\omega_{1}}-
\frac{J_{\alpha}(\Gamma_{2})-J_{\alpha}(-\Gamma_{2})}{2}
\mbox{e}^{it\alpha\omega_{2}}
\right\}  \nonumber \\
&\equiv& \mbox{e}^{-it\omega x^{2}}B_{0},     \nonumber  \\
C&=&\mbox{e}^{-it\omega x^{2}}
\frac{\Delta}{2}\bra{n}\mbox{e}^{x(a^{\dagger}-a)}\ket{n}\sum_{\alpha\neq 0}
\left\{
\frac{J_{\alpha}(\Gamma_{1})-J_{\alpha}(-\Gamma_{1})}{2}
\mbox{e}^{it\alpha\omega_{1}}+
\frac{J_{\alpha}(\Gamma_{2})-J_{\alpha}(-\Gamma_{2})}{2}
\mbox{e}^{it\alpha\omega_{2}}
\right\}  \nonumber \\
&\equiv& \mbox{e}^{-it\omega x^{2}}C_{0},  \nonumber \\
D&=&\mbox{e}^{-it\omega x^{2}}
\frac{\Delta}{2}\bra{n}\mbox{e}^{x(a^{\dagger}-a)}\ket{n}\sum_{\alpha\neq 0}
\left\{
\frac{J_{\alpha}(\Gamma_{1})+J_{\alpha}(-\Gamma_{1})}{2}
\mbox{e}^{it\alpha\omega_{1}}-
\frac{J_{\alpha}(\Gamma_{2})+J_{\alpha}(-\Gamma_{2})}{2}
\mbox{e}^{it\alpha\omega_{2}}
\right\}  \nonumber \\
&\equiv& \mbox{e}^{-it\omega x^{2}}D_{0}. 
\end{eqnarray}
We note that ${\bar A}_{0}=A_{0},\ {\bar D}_{0}=D_{0}$ and 
${\bar B}_{0}=-B_{0},\ {\bar C}_{0}=-C_{0}$ (due to the fact 
$J_{-\alpha}(x)=J_{\alpha}(-x)$). 

The above matrix equation decomposes into two parts 
\begin{eqnarray}
i\frac{d}{dt}
\left(
  \begin{array}{c}
    c_{n,1}\\
    c_{n,3}
  \end{array}
\right)
=      
&&\left(
  \begin{array}{cc}
    u_{n,11}& u_{n,13}   \\
    u_{n,31}& u_{n,33}   
  \end{array}
\right)^{-1}
\left(
  \begin{array}{cc}
           0&      A     \\
    {\bar A}&      0     \\
  \end{array}
\right)
\left(
  \begin{array}{cc}
    u_{n,11}& u_{n,13}  \\
    u_{n,31}& u_{n,33}  
  \end{array}
\right)
\left(
  \begin{array}{c}
    c_{n,1}\\
    c_{n,3}
  \end{array}
\right)
+   \nonumber \\
&&
\left(
  \begin{array}{cc}
    u_{n,11}& u_{n,13}   \\
    u_{n,31}& u_{n,33}   
  \end{array}
\right)^{-1}
\left(
  \begin{array}{cc}
           0&     B    \\
    {\bar C}&     0    \\
  \end{array}
\right)
\left(
  \begin{array}{cc}
    u_{n,22}& u_{n,24}   \\
    u_{n,42}& u_{n,44}   
  \end{array}
\right)
\left(
  \begin{array}{c}
    c_{n,2}\\
    c_{n,4}
  \end{array}
\right),    \nonumber \\
i\frac{d}{dt}
\left(
  \begin{array}{c}
    c_{n,2}\\
    c_{n,4}
  \end{array}
\right)
=  
&&\left(
  \begin{array}{cc}
    u_{n,22}& u_{n,24}   \\
    u_{n,42}& u_{n,44}   
  \end{array}
\right)^{-1}
\left(
  \begin{array}{cc}
           0&     C    \\
     {\bar B}&     0    \\
  \end{array}
\right)
\left(
  \begin{array}{cc}
    u_{n,11}& u_{n,13}   \\
    u_{n,31}& u_{n,33}   
  \end{array}
\right)
\left(
  \begin{array}{c}
    c_{n,1}\\
    c_{n,3}
  \end{array}
\right)+   \nonumber \\
&&\left(
  \begin{array}{cc}
    u_{n,22}& u_{n,24}   \\
    u_{n,42}& u_{n,44}   
  \end{array}
\right)^{-1}
\left(
  \begin{array}{cc}
            0&    D    \\
     {\bar D}&    0    \\
  \end{array}
\right)
\left(
  \begin{array}{cc}
    u_{n,22}& u_{n,24}   \\
    u_{n,42}& u_{n,44}   
  \end{array}
\right)
\left(
  \begin{array}{c}
    c_{n,2}\\
    c_{n,4}
  \end{array}
\right).    \nonumber 
\end{eqnarray}

By the way, from (\ref{eq:special-unitary-1}), (\ref{eq:special-unitary-2}) 
and (\ref{eq:coefficients}) 
\begin{eqnarray}
&&i\frac{d}{dt}
\left(
  \begin{array}{c}
    c_{n,1}\\
    c_{n,3}
  \end{array}
\right)
=                 \nonumber \\
&&A_{0}
\mbox{exp}\left\{it
\left(
  \begin{array}{cc}
    0& E_{\Delta,n,+} \\
    E_{\Delta,n,+}& \omega x^{2}
  \end{array}
\right)
\right\}
\left(
  \begin{array}{cc}
           0&      1     \\
           1&      0     \\
  \end{array}
\right)
\mbox{exp}\left\{-it
\left(
  \begin{array}{cc}
    0& E_{\Delta,n,+} \\
    E_{\Delta,n,+}& \omega x^{2}
  \end{array}
\right)
\right\}
\left(
  \begin{array}{c}
    c_{n,1}\\
    c_{n,3}
  \end{array}
\right)+           \nonumber \\  
&&
\mbox{exp}\left\{it
\left(
  \begin{array}{cc}
    0& E_{\Delta,n,+} \\
    E_{\Delta,n,+}& \omega x^{2}
  \end{array}
\right)
\right\}
\left(
  \begin{array}{cc}
               0&   B_{0}  \\
    {\bar C_{0}}&    0     \\
  \end{array}
\right)
\mbox{exp}\left\{-it
\left(
  \begin{array}{cc}
    0& E_{\Delta,n,-} \\
    E_{\Delta,n,-}& \omega x^{2}
  \end{array}
\right)
\right\}
\left(
  \begin{array}{c}
    c_{n,2}\\
    c_{n,4}
  \end{array}
\right),      \nonumber \\
&&{}          \\   
&&i\frac{d}{dt}
\left(
  \begin{array}{c}
    c_{n,2}\\
    c_{n,4}
  \end{array}
\right)
=               \nonumber \\
&&
\mbox{exp}\left\{it
\left(
  \begin{array}{cc}
    0& E_{\Delta,n,-} \\
    E_{\Delta,n,-}& \omega x^{2}
  \end{array}
\right)
\right\}
\left(
  \begin{array}{cc}
               0&   C_{0}  \\
    {\bar B_{0}}&    0     \\
  \end{array}
\right)
\mbox{exp}\left\{-it
\left(
  \begin{array}{cc}
    0& E_{\Delta,n,+} \\
    E_{\Delta,n,+}& \omega x^{2}
  \end{array}
\right)
\right\}
\left(
  \begin{array}{c}
    c_{n,1}\\
    c_{n,3}
  \end{array}
\right)+     \nonumber  \\
&&D_{0}
\mbox{exp}\left\{it
\left(
  \begin{array}{cc}
    0& E_{\Delta,n,-} \\
    E_{\Delta,n,-}& \omega x^{2}
  \end{array}
\right)
\right\}
\left(
  \begin{array}{cc}
           0&      1     \\
           1&      0     
  \end{array}
\right)
\mbox{exp}\left\{-it
\left(
  \begin{array}{cc}
    0& E_{\Delta,n,-}              \\
    E_{\Delta,n,-}& \omega x^{2}
  \end{array}
\right)
\right\}
\left(
  \begin{array}{c}
    c_{n,2}\\
    c_{n,4}
  \end{array}
\right).          \nonumber \\
&&{} 
\end{eqnarray}

At this stage we can set several resonance conditions and obtain 
corresponding equations and solutions by making use of the rotating wave 
approximation like in \cite{KF7}, \cite{KF8}. For example, 
let us consider the term 
\[
\mbox{exp}\left\{it
\left(
  \begin{array}{cc}
    0& E_{\Delta,n,+} \\
    E_{\Delta,n,+}& \omega x^{2}
  \end{array}
\right)
\right\}
\left(
  \begin{array}{cc}
                 0&   B_{0}    \\
      {\bar C_{0}}&      0     
  \end{array}
\right)
\mbox{exp}\left\{-it
\left(
  \begin{array}{cc}
    0& E_{\Delta,n,-} \\
    E_{\Delta,n,-}& \omega x^{2}
  \end{array}
\right)
\right\},
\]
then from (\ref{eq:appendix-Q}) in Appendix 
\[
P_{n,+}
\left(
  \begin{array}{cc}
    \mbox{e}^{it\mu_{n,+}}&                         \\
                          & \mbox{e}^{it\nu_{n,+}}
  \end{array}
\right)
P_{n,+}^{-1}
\left(
  \begin{array}{cc}
                 0&   B_{0}    \\
      {\bar C_{0}}&      0     
  \end{array}
\right)
P_{n,-}
\left(
  \begin{array}{cc}     
    \mbox{e}^{-it\mu_{n,-}}&                         \\
                           & \mbox{e}^{-it\nu_{n,-}}
  \end{array}
\right)
P_{n,-}^{-1}.
\]
Here for example, 
\[
\mbox{e}^{it(\mu_{n,+}-\mu_{n,-})}
P_{n,+}
\left(
  \begin{array}{cc}
    1&                         \\
     & \mbox{e}^{it(\nu_{n,+}-\mu_{n,+})}
  \end{array}
\right)
P_{n,+}^{-1}
\left(
  \begin{array}{cc}
                 0&   B_{0}    \\
      {\bar C_{0}}&      0     
  \end{array}
\right)
P_{n,-}
\left(
  \begin{array}{cc}     
     1&                           \\
      & \mbox{e}^{-it(\nu_{n,-}-\mu_{n,-})}
  \end{array}
\right)
P_{n,-}^{-1}
\]
for 
\begin{eqnarray}
B_{0}&=&
\frac{\Delta}{2}\bra{n}\mbox{e}^{x(a^{\dagger}-a)}\ket{n}\sum_{\alpha\neq 0}
\left\{
\frac{J_{\alpha}(\Gamma_{1})-J_{\alpha}(-\Gamma_{1})}{2}
\mbox{e}^{it\alpha\omega_{1}}-
\frac{J_{\alpha}(\Gamma_{2})-J_{\alpha}(-\Gamma_{2})}{2}
\mbox{e}^{it\alpha\omega_{2}}
\right\},  \nonumber \\
{\bar C}_{0}&=&
\frac{\Delta}{2}\bra{n}\mbox{e}^{x(a^{\dagger}-a)}\ket{n}\sum_{\alpha\neq 0}
\left\{
\frac{J_{\alpha}(-\Gamma_{1})-J_{\alpha}(\Gamma_{1})}{2}
\mbox{e}^{it\alpha\omega_{1}}-
\frac{J_{\alpha}(\Gamma_{2})-J_{\alpha}(-\Gamma_{2})}{2}
\mbox{e}^{it\alpha\omega_{2}}
\right\}.  \nonumber 
\end{eqnarray}

Now we set the resonance condition 
\begin{equation}
\label{eq:resonance condition}
\alpha\omega_{2}+\mu_{n,+}-\mu_{n,-}=0
\Longleftrightarrow 
(-\alpha)\omega_{2}+\mu_{n,-}-\mu_{n,+}=0
\end{equation}
for some $\alpha\neq 0$, then we can neglect the remaining terms by 
the rotating wave approximation. That is, we obtain the time independent 
matrix
\[
-\frac{\Delta}{2}\bra{n}\mbox{e}^{x(a^{\dagger}-a)}\ket{n}
\frac{J_{\alpha}(\Gamma_{2})-J_{\alpha}(-\Gamma_{2})}{2}
P_{n,+}
\left(
  \begin{array}{cc}
    1&    \\
     & 0
  \end{array}
\right)
P_{n,+}^{-1}
\left(
  \begin{array}{cc}
                 0&  1    \\
                 1&  0    
  \end{array}
\right)
P_{n,-}
\left(
  \begin{array}{cc}     
     1&     \\
      & 0
  \end{array}
\right)
P_{n,-}^{-1}, 
\]
so we have 
\begin{eqnarray}
i\frac{d}{dt}
\left(
  \begin{array}{c}
    c_{n,1}\\
    c_{n,3}
  \end{array}
\right)
&=&        
-\frac{\Delta}{2}\bra{n}\mbox{e}^{x(a^{\dagger}-a)}\ket{n}
\frac{J_{\alpha}(\Gamma_{2})-J_{\alpha}(-\Gamma_{2})}{2}
\times     \nonumber \\
&&P_{n,+}
\left(
  \begin{array}{cc}
    1&    \\
     & 0
  \end{array}
\right)
P_{n,+}^{-1}
\left(
  \begin{array}{cc}
                 0&  1    \\
                 1&  0    
  \end{array}
\right)
P_{n,-}
\left(
  \begin{array}{cc}     
     1&     \\
      & 0
  \end{array}
\right)
P_{n,-}^{-1}
\left(
  \begin{array}{c}
    c_{n,2}\\
    c_{n,4}
  \end{array}
\right),      \nonumber   \\
i\frac{d}{dt}
\left(
  \begin{array}{c}
    c_{n,2}\\
    c_{n,4}
  \end{array}
\right)
&=&        
-\frac{\Delta}{2}\bra{n}\mbox{e}^{x(a^{\dagger}-a)}\ket{n}
\frac{J_{\alpha}(\Gamma_{2})-J_{\alpha}(-\Gamma_{2})}{2}
\times     \nonumber \\
&&P_{n,-}
\left(
  \begin{array}{cc}
    1&    \\
     & 0
  \end{array}
\right)
P_{n,-}^{-1}
\left(
  \begin{array}{cc}
                 0&  1    \\
                 1&  0    
  \end{array}
\right)
P_{n,+}
\left(
  \begin{array}{cc}     
     1&     \\
      & 0
  \end{array}
\right)
P_{n,+}^{-1}
\left(
  \begin{array}{c}
    c_{n,1}\\
    c_{n,3}
  \end{array}
\right),        \nonumber
\end{eqnarray}
where we have used the relation 
\[
\frac{J_{-\alpha}(\Gamma_{2})-J_{-\alpha}(-\Gamma_{2})}{2}
=
-\frac{J_{\alpha}(\Gamma_{2})-J_{\alpha}(-\Gamma_{2})}{2}.
\]
After some algebras we obtain 
\begin{eqnarray}
i\frac{d}{dt}
\left(
  \begin{array}{c}
     c_{n,1} \\
     c_{n,3}
  \end{array}
\right)&&
=       
-\frac{\Delta}{2}\bra{n}\mbox{e}^{x(a^{\dagger}-a)}\ket{n}
\frac{J_{\alpha}(\Gamma_{2})-J_{\alpha}(-\Gamma_{2})}{2}
\times     \nonumber \\
&&\frac{E_{\Delta,n,+}\mu_{n,-}+E_{\Delta,n,-}\mu_{n,+}}
{(E_{\Delta,n,+}^{2}+\mu_{n,+}^{2})(E_{\Delta,n,-}^{2}+\mu_{n,-}^{2})}
\left(
  \begin{array}{cc}     
     E_{\Delta,n,+}E_{\Delta,n,-}&  E_{\Delta,n,+}\mu_{n,-}   \\
     \mu_{n,+}E_{\Delta,n,-}& \mu_{n,+}\mu_{n,-}
  \end{array}
\right)
\left(
  \begin{array}{c}
    c_{n,2}\\
    c_{n,4}
  \end{array}
\right),           \\
i\frac{d}{dt}
\left(
  \begin{array}{c}
     c_{n,2} \\
     c_{n,4}
  \end{array}
\right)&&
=        
-\frac{\Delta}{2}\bra{n}\mbox{e}^{x(a^{\dagger}-a)}\ket{n}
\frac{J_{\alpha}(\Gamma_{2})-J_{\alpha}(-\Gamma_{2})}{2}
\times     \nonumber \\
&&\frac{E_{\Delta,n,+}\mu_{n,-}+E_{\Delta,n,-}\mu_{n,+}}
{(E_{\Delta,n,+}^{2}+\mu_{n,+}^{2})(E_{\Delta,n,-}^{2}+\mu_{n,-}^{2})}
\left(
  \begin{array}{cc}     
     E_{\Delta,n,+}E_{\Delta,n,-}&  \mu_{n,+}E_{\Delta,n,-}   \\
     E_{\Delta,n,+}\mu_{n,-}& \mu_{n,+}\mu_{n,-}
  \end{array}
\right)
\left(
  \begin{array}{c}
    c_{n,1}\\
    c_{n,3}
  \end{array}
\right).
\end{eqnarray}
from the appendix. 

If we prepare the vector notations 
\[
{\bf c}_{od}=
\left(
  \begin{array}{c}
    c_{n,1}\\
    c_{n,3}
  \end{array}
\right),
\quad 
{\bf c}_{ev}=
\left(
  \begin{array}{c}
    c_{n,2}\\
    c_{n,4}
  \end{array}
\right),
\]
then the equation above can be written as
\begin{equation}
i\frac{d}{dt}
\left(
  \begin{array}{c}
    {\bf c}_{od}\\
    {\bf c}_{ev}
  \end{array}
\right)
=       
-\frac{{\cal R}}{2}
\left(
  \begin{array}{cc}
         & K  \\
    K^{T}& 
  \end{array}
\right)
\left(
  \begin{array}{c}
    {\bf c}_{od}\\
    {\bf c}_{ev}
  \end{array}
\right)
\end{equation}
with 
\[
{\cal R}=
\frac{\Delta}{2}\bra{n}\mbox{e}^{x(a^{\dagger}-a)}\ket{n}
\frac{J_{\alpha}(\Gamma_{2})-J_{\alpha}(-\Gamma_{2})}{2}
\frac{E_{\Delta,n,+}\mu_{n,-}+E_{\Delta,n,-}\mu_{n,+}}
{\sqrt{(E_{\Delta,n,+}^{2}+\mu_{n,+}^{2})(E_{\Delta,n,-}^{2}+\mu_{n,-}^{2})}}
\]
and
\[
K=
\frac{1}
{\sqrt{(E_{\Delta,n,+}^{2}+\mu_{n,+}^{2})(E_{\Delta,n,-}^{2}+\mu_{n,-}^{2})}}
\left(
  \begin{array}{cc}     
     E_{\Delta,n,+}E_{\Delta,n,-}& E_{\Delta,n,+}\mu_{n,-}  \\
     \mu_{n,+}E_{\Delta,n,-}& \mu_{n,+}\mu_{n,-}
  \end{array}
\right).
\]
The (formal) solution is easily obtained to become 
\begin{equation}
\label{eq:solution}
\left(
  \begin{array}{c}
    {\bf c}_{od}(t)\\
    {\bf c}_{ev}(t)
  \end{array}
\right)
= 
\mbox{exp}
\left\{      
\frac{i{\cal R}t}{2}
\left(
  \begin{array}{cc}
         & K  \\
    K^{T}& 
  \end{array}
\right)
\right\}
\left(
  \begin{array}{c}
    {\bf c}_{od}(0)\\
    {\bf c}_{ev}(0)
  \end{array}
\right).
\end{equation}

Next let us calculate the unitary transformation of (\ref{eq:solution}). 
It is easy to see 
\begin{eqnarray}
\label{eq:kei}
K&=&
\frac{1}
{\sqrt{(E_{\Delta,n,+}^{2}+\mu_{n,+}^{2})(E_{\Delta,n,-}^{2}+\mu_{n,-}^{2})}}
\left(
  \begin{array}{c}   
    E_{\Delta,n,+} \\
    \mu_{n,+}
  \end{array}
\right)
{\bf \Bigl(}
    E_{\Delta,n,-},\ \mu_{n,-}
{\bf \Bigr)},  \\
\label{eq:kei-t}
K^{T}&=&
\frac{1}
{\sqrt{(E_{\Delta,n,+}^{2}+\mu_{n,+}^{2})(E_{\Delta,n,-}^{2}+\mu_{n,-}^{2})}}
\left(
  \begin{array}{c}   
    E_{\Delta,n,-} \\
    \mu_{n,-}
  \end{array}
\right)
{\bf \Bigl(}
    E_{\Delta,n,+},\ \mu_{n,+}
{\bf \Bigr)}, 
\end{eqnarray}
so 
\begin{eqnarray}
K^{T}K
&=&\frac{1}{E_{\Delta,n,-}^{2}+\mu_{n,-}^{2}}
\left(
  \begin{array}{c}   
    E_{\Delta,n,-} \\
    \mu_{n,-}
  \end{array}
\right)
{\bf \Bigl(}
    E_{\Delta,n,-},\ \mu_{n,-}
{\bf \Bigr)}      \nonumber \\
&=&
\left(
  \begin{array}{c}   
    E_{\Delta,n,-} \\
    \mu_{n,-}
  \end{array}
\right)
\left(E_{\Delta,n,-}^{2}+\mu_{n,-}^{2}\right)^{-1}
{\bf \Bigl(}
    E_{\Delta,n,-},\ \mu_{n,-}
{\bf \Bigr)},    \nonumber \\
KK^{T}
&=&\frac{1}{E_{\Delta,n,+}^{2}+\mu_{n,+}^{2}}
\left(
  \begin{array}{c}   
    E_{\Delta,n,+} \\
    \mu_{n,+}
  \end{array}
\right)
{\bf \Bigl(}
    E_{\Delta,n,+},\ \mu_{n,+}
{\bf \Bigr)}       \nonumber \\
&=&
\left(
  \begin{array}{c}   
    E_{\Delta,n,+} \\
    \mu_{n,+}
  \end{array}
\right)
\left(E_{\Delta,n,+}^{2}+\mu_{n,+}^{2}\right)^{-1}
{\bf \Bigl(}
    E_{\Delta,n,+},\ \mu_{n,+}
{\bf \Bigr)}.    \nonumber 
\end{eqnarray}
Namely, $K^{T}K$ and $KK^{T}$ are just projection matrices and moreover 
satisfy 
\[
(KK^{T})^{n}=KK^{T},\quad (K^{T}K)^{n}=K^{T}K,\quad 
KK^{T}K=K,\quad K^{T}KK^{T}=K^{T} 
\]
for $n\geq 1$. Therefore after long algebras using the relations above 
we have 
\begin{eqnarray}
&&\mbox{exp}
\left\{      
\frac{i{\cal R}t}{2}
\left(
  \begin{array}{cc}
         & K  \\
    K^{T}& 
  \end{array}
\right)
\right\}
=           \nonumber \\
&&\left(
  \begin{array}{cc}
     {\bf 1}-KK^{T}+\mbox{cos}\left(\frac{{\cal R}t}{2}\right)KK^{T} &
       i\mbox{sin}\left(\frac{{\cal R}t}{2}\right)K  \\
    i\mbox{sin}\left(\frac{{\cal R}t}{2}\right)K^{T} & 
       {\bf 1}-K^{T}K+\mbox{cos}\left(\frac{{\cal R}t}{2}\right)K^{T}K
  \end{array}
\right),
\end{eqnarray}
so the solution is explicitly 
\begin{eqnarray}
\label{eq:explicit-solution}
&&
\left(
  \begin{array}{c}
    {\bf c}_{od}(t)\\
    {\bf c}_{ev}(t)
  \end{array}
\right)
= 
\left(
  \begin{array}{cc}
     {\bf 1}-KK^{T}+\mbox{cos}\left(\frac{{\cal R}t}{2}\right)KK^{T} &
       i\mbox{sin}\left(\frac{{\cal R}t}{2}\right)K  \\
    i\mbox{sin}\left(\frac{{\cal R}t}{2}\right)K^{T} & 
       {\bf 1}-K^{T}K+\mbox{cos}\left(\frac{{\cal R}t}{2}\right)K^{T}K
  \end{array}
\right)
\left(
  \begin{array}{c}
    {\bf c}_{od}(0)\\
    {\bf c}_{ev}(0)
  \end{array}
\right)       \nonumber \\
&&{}
\end{eqnarray}
with (\ref{eq:kei}) and (\ref{eq:kei-t}).

\par \noindent
This is one of unitary transformations that we are looking for in Quantum 
Computation. We can also obtain another ones by setting different resonance 
conditions like (\ref{eq:resonance condition}) (we leave them to the readers).

\par \vspace{5mm} 
Lastly let us summarize our result. By (\ref{eq:general-hamiltonian}) 
the Hamiltonian becomes 
\begin{equation}
H_{L}
=
\sum_{j=1}^{m}
\left\{
\frac{\Delta}{2}\sigma_{3}^{(j)}+ 
g_{2}\mbox{cos}(\omega_{j}t+\phi_{j})\sigma_{1}^{(j)}
\right\}
\otimes {\bf 1}_{L}
\end{equation}
if there is no photon interaction, so the dynamics of each qubit space 
is independently determined by 
\begin{equation}
H_{Lj}
=
\frac{\Delta}{2}\sigma_{3}^{(j)}+ 
g_{2}\mbox{cos}(\omega_{j}t+\phi_{j})\sigma_{1}^{(j)}. 
\end{equation}
Therefore the total space of $m$--qubits is just 
\[
\fukuso^{2}\otimes \cdots \otimes \fukuso^{2}\otimes \cdots \otimes \fukuso^{2}
\quad \mbox{where}\quad 
\fukuso^{2}=\mbox{Vect}_{\fukuso}\{\ket{0},\ \ket{1}\}, 
\]
and unitary transformations of each qubit space are obtained by manipulating 
laser fields. 

\par \noindent
However to solve the Schr{\"o}dinger equation (neglecting the suffix)
\[
i\frac{d}{dt}\Psi=
\left\{
\frac{\Delta}{2}\sigma_{3}+ 
g_{2}\mbox{cos}(\omega t+\phi)\sigma_{1}
\right\}
\Psi
\]
is not so easy, see for example \cite{MFr4}. 

When $m$ = 2, the interaction (driving by the photon) between two qubits is 
given by unitary transformations like (\ref{eq:explicit-solution}) 
$\cdots$ controlled unitary gates including the controlled NOT. 

\vspace{10mm}
%%%%%%%%%%%%%%%%%%%%%%%%%%%%%%%%%%%%%%%%%%%%%%%%%%%%%%%%%%%%%%%%%%%%
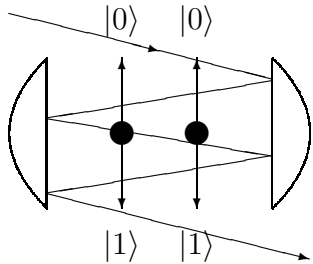
\begin{center}
\setlength{\unitlength}{1mm} 
\begin{picture}(80,40)(0,-20)
\bezier{200}(20,0)(10,10)(20,20)
%
%\put(15,26){\line(4,-1){35}}
%\put(50,17){\line(-6,-1){30}}
%\put(20,12){\line(6,-1){30}}
%\put(50, 7){\line(-6,-1){30}}
%\put(20, 2){\line(4,-1){35}}
%
\put(15,26){\vector(4,-1){20}}
\put(35,21){\line(4,-1){15}}
\put(50,17){\line(-6,-1){30}}
\put(20,12){\line(6,-1){30}}
\put(50, 7){\line(-6,-1){30}}
\put(20, 2){\vector(4,-1){35}}
\put(20,0){\line(0,1){20}}
\put(20,20){\makebox(20,10)[c]{$|0\rangle$}}
\put(30,10){\vector(0,1){10}}
\put(30,10){\vector(0,-1){10}}
\put(20,-10){\makebox(20,10)[c]{$|1\rangle$}}
\put(30,10){\circle*{3}}
\put(30,20){\makebox(20,10)[c]{$|0\rangle$}}
\put(40,10){\vector(0,1){10}}
\put(40,10){\vector(0,-1){10}}
\put(30,-10){\makebox(20,10)[c]{$|1\rangle$}}
\put(40,10){\circle*{3}}
\bezier{200}(50,0)(60,10)(50,20)
\put(50,0){\line(0,1){20}}
\end{picture}
\end{center}
%%%%%%%%%%%%%%%%%%%%%%%%%%%%%%%%%%%%%%%%%%%%%%%%%%%%%%%%%%%%%%%%%%%%
\vspace{-10mm} 

\par \noindent
This is just our scenario of Quantum Computation in the strong coupling 
regime.

\vspace{10mm}
\begin{flushleft}
{\sl {\Large Case of m=3}}\ {\Large :}
\end{flushleft}
We present a very important 

\par \noindent
{\bf Problem : }Let us consider three atoms in a cavity. How can we 
construct C-NOT (or C-Unitary) operations for any two atoms among them ? 

\par \noindent
See the following pictures : 

\vspace{10mm}
%%%%%%%%%%%%%%%%%%%%%%%%%%%%%%%%%%%%%%%%%%%%%%%%%%%%%%%%%%%%%%%%%%%%
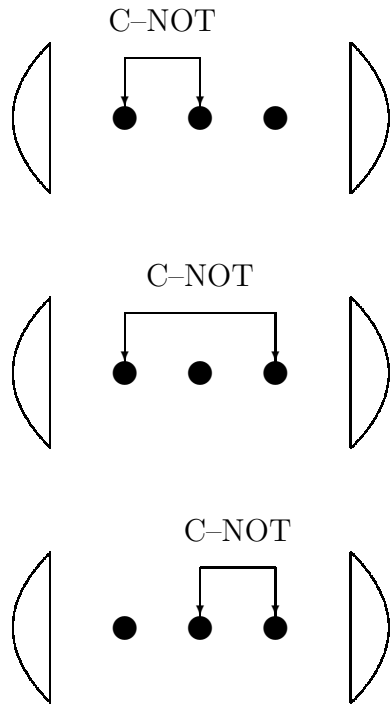
\begin{center}
\setlength{\unitlength}{1mm} 
\begin{picture}(90,40)(0,-20)
\bezier{200}(20,0)(10,10)(20,20)
\put(20,0){\line(0,1){20}}
\put(30,18){\vector(0,-1){6.5}}
\put(30,10){\circle*{3}}
\put(40,18){\vector(0,-1){6.5}}
\put(40,10){\circle*{3}}
%\put(50,18){\vector(0,-1){6.5}}
\put(50,10){\circle*{3}}
\put(25,18){\makebox(20,10)[c]{C--NOT}}
\put(30,18){\line(1,0){10}}
\bezier{200}(60,0)(70,10)(60,20)
\put(60,0){\line(0,1){20}}
\end{picture}
\end{center}
%%%%%%%%%%%%%%%%%%%%%%%%%%%%%%%%%%%%%%%%%%%%%%%%%%%%%%%%%%%%%%%%%%%%
%
\vspace{-15mm}
%%%%%%%%%%%%%%%%%%%%%%%%%%%%%%%%%%%%%%%%%%%%%%%%%%%%%%%%%%%%%%%%%%%%
\begin{center}
\setlength{\unitlength}{1mm} 
\begin{picture}(90,40)(0,-20)
\bezier{200}(20,0)(10,10)(20,20)
\put(20,0){\line(0,1){20}}
\put(30,18){\vector(0,-1){6.5}}
\put(30,10){\circle*{3}}
\put(40,10){\circle*{3}}
\put(50,18){\vector(0,-1){6.5}}
\put(50,10){\circle*{3}}
\put(30,18){\makebox(20,10)[c]{C--NOT}}
\put(30,18){\line(1,0){20}}
\bezier{200}(60,0)(70,10)(60,20)
\put(60,0){\line(0,1){20}}
\end{picture}
\end{center}
%%%%%%%%%%%%%%%%%%%%%%%%%%%%%%%%%%%%%%%%%%%%%%%%%%%%%%%%%%%%%%%%%%%%
%
\vspace{-15mm}
%%%%%%%%%%%%%%%%%%%%%%%%%%%%%%%%%%%%%%%%%%%%%%%%%%%%%%%%%%%%%%%%%%%%
\begin{center}
\setlength{\unitlength}{1mm} 
\begin{picture}(90,40)(0,-20)
\bezier{200}(20,0)(10,10)(20,20)
\put(20,0){\line(0,1){20}}
\put(30,10){\circle*{3}}
\put(40,18){\vector(0,-1){6.5}}
\put(40,10){\circle*{3}}
\put(50,18){\vector(0,-1){6.5}}
\put(50,10){\circle*{3}}
\put(25,18){\makebox(40,10)[c]{C--NOT}}
\put(50,18){\line(-1,0){10}}
\bezier{200}(60,0)(70,10)(60,20)
\put(60,0){\line(0,1){20}}
\end{picture}
\end{center}
%%%%%%%%%%%%%%%%%%%%%%%%%%%%%%%%%%%%%%%%%%%%%%%%%%%%%%%%%%%%%%%%%%%%
%

\vspace{-15mm} 
These constructions are very crucial in realizing quantum logic gates, 
however we have not seen such constructions in any references. 
We will attack this problem in a forthcoming paper.

\par \vspace{5mm}
In this paper we treated the two--atoms case in a cavity QED and 
constructed unitary transformations by making use of the rotating wave 
approximation under new resonance conditions containing the Bessel 
functions.

These will be applied to construct several quantum logic gates in Quantum 
Computation. 
Moreover we would like to treat a general case, which is at the present not 
easy due to some technical reasons. 

By the way,  according to increase of the number of atoms (we are expecting 
at least $m=100$ in the realistic quantum computation) we meet a very severe 
problem called Decoherence, see for example \cite{MFr3} and its references. 
We don't know how to control this. 

\par \noindent
One way protecting against this may be to deal with N--level system (then 
we can reduce the number of atoms in a cavity). 
A generalization of the model to N--level system (see for example \cite{KF3}, 
\cite{KF6}, \cite{FHKW}, \cite{KuF}) is now under consideration and will be 
published in a separate paper.

\noindent \vspace{3mm}

{\it Acknowledgment.}
The author wishes to thank Marco Frasca for his important suggestions. 
He also wishes to thank the graduate students Kyoko Higashida, 
Ryosuke Kato and Yukako Wada for some help.

\vspace{10mm}
\begin{center}
\begin{Large}
\noindent{\bfseries Appendix : Some Useful Formulas}
\end{Large}
\end{center}
\par \vspace{5mm} \noindent
In this appendix we solve the following equation 
\begin{equation}
i\frac{d}{dt}\psi=\alpha H\psi, 
\end{equation}
where $\alpha$ is a constant and 
\begin{equation}
H=\mbox{e}^{-i\theta t}\ket{1}\bra{-1}+\mbox{e}^{i\theta t}\ket{-1}\bra{1}
\quad \mbox{and}\quad \psi=a(t)\ket{1}+b(t)\ket{-1}. 
\end{equation}
Then it is easy to get a matrix equation on $\{a,\ b\}$ 
\begin{equation}
\label{eq:2-matrix-equation}
i\frac{d}{dt}
\left(
  \begin{array}{c}
    a \\
    b
  \end{array}
\right)
=
\left(
  \begin{array}{cc}
    0                         & \alpha\mbox{e}^{-i\theta t} \\
    \alpha\mbox{e}^{i\theta t}& 0
  \end{array}
\right)
\left(
  \begin{array}{c}
    a \\
    b
  \end{array}
\right)\Longleftrightarrow 
i\frac{d}{dt}{\tilde \psi}={\tilde H}{\tilde \psi}.
\end{equation}
Noting the decomposition 
\[
\left(
  \begin{array}{cc}
    0                         & \alpha\mbox{e}^{-i\theta t} \\
    \alpha\mbox{e}^{i\theta t}& 0
  \end{array}
\right)
=
\left(
  \begin{array}{cc}
    1& \\
     &\mbox{e}^{i\theta t}
  \end{array}
\right)
\left(
  \begin{array}{cc}
    0& \alpha \\
    \alpha& 0
  \end{array}
\right)
\left(
  \begin{array}{cc}
    1& \\
     &\mbox{e}^{-i\theta t}
  \end{array}
\right),
\]
then the equation becomes 
\[
i\frac{d}{dt}{\hat \psi}
=
\left(
  \begin{array}{cc}
    0     & \alpha \\
    \alpha& \theta
  \end{array}
\right)
{\hat \psi}
\qquad \mbox{for} \quad 
{\hat \psi}\equiv 
\left(
  \begin{array}{cc}
    1&                      \\
     &\mbox{e}^{-i\theta t}
  \end{array}
\right)
{\tilde \psi}.
\]
The solution is easily obtained to become 
\begin{equation}
\left(
  \begin{array}{c}
    a \\
    b
  \end{array}
\right)
=U(t)
\left(
  \begin{array}{c}
    a_{0} \\
    b_{0}
  \end{array}
\right)
\end{equation}
where $(a_{0},b_{0})^{T}$ is a constant vector and 
\begin{equation}
U(t)=
\left(
  \begin{array}{cc}
    1& \\
     & \mbox{e}^{i\theta t}
  \end{array}
\right)
\mbox{exp}\left\{-it
\left(
  \begin{array}{cc}
    0     & \alpha \\
    \alpha& \theta
  \end{array}
\right)
\right\}\Longrightarrow i\frac{d}{dt}U={\tilde H}U.
\end{equation}

\par \noindent
If we set 
\begin{equation}
U(t)=
\left(
  \begin{array}{cc}
    u_{11}& u_{12}\\
    u_{21}& u_{22}
  \end{array}
\right)
\end{equation}
($2$ corresponds to $-1$) then $\psi$ above can be written as 
\begin{equation}
\psi=(u_{11}a_{0}+u_{12}b_{0})\ket{1}+(u_{21}a_{0}+u_{22}b_{0})\ket{-1}
\end{equation}
with constants $\{a_{0},\ b_{0}\}$. 

\par \noindent 
In the method of constant variation in the text we change like 
$a_{0}\longrightarrow a_{0}(t)$ and $b_{0}\longrightarrow b_{0}(t)$. 

Let us make some comments. For 
\begin{equation}
A=
\left(
  \begin{array}{cc}
    0     & \alpha  \\
    \alpha& \theta
  \end{array}
\right)
\end{equation}
we can easily diagonalize $A$ as follows : 
\begin{equation}
A=
\left(
  \begin{array}{cc}
    \frac{\alpha}{\sqrt{\alpha^2+\mu^2}}& 
    \frac{\alpha}{\sqrt{\alpha^2+\nu^2}}\\
    \frac{\mu}{\sqrt{\alpha^2+\mu^2}}& 
    \frac{\nu}{\sqrt{\alpha^2+\nu^2}}
  \end{array}
\right)
\left(
  \begin{array}{cc}
    \mu& \\
      & \nu
  \end{array}
\right)
\left(
  \begin{array}{cc}
    \frac{\alpha}{\sqrt{\alpha^2+\mu^2}}& 
    \frac{\alpha}{\sqrt{\alpha^2+\nu^2}}\\
    \frac{\mu}{\sqrt{\alpha^2+\mu^2}}& 
    \frac{\nu}{\sqrt{\alpha^2+\nu^2}}
  \end{array}
\right)^{-1}
\end{equation}
where 
\[
\mu=\frac{1}{2}(\theta+\sqrt{\theta^{2}+4\alpha^{2}}),
\quad 
\nu=\frac{1}{2}(\theta-\sqrt{\theta^{2}+4\alpha^{2}}).
\]
Therefore we obtain 
\begin{equation}
\label{eq:appendix-Q}
Q(t)\equiv \mbox{e}^{-itA}= 
\left(
  \begin{array}{cc}
    \frac{\alpha}{\sqrt{\alpha^2+\mu^2}}& 
    \frac{\alpha}{\sqrt{\alpha^2+\nu^2}}\\
    \frac{\mu}{\sqrt{\alpha^2+\mu^2}}& 
    \frac{\nu}{\sqrt{\alpha^2+\nu^2}}
  \end{array}
\right)
\left(
  \begin{array}{cc}
    \mbox{e}^{-it \mu}&                    \\
                      & \mbox{e}^{-it \nu}
  \end{array}
\right)
\left(
  \begin{array}{cc}
    \frac{\alpha}{\sqrt{\alpha^2+\mu^2}}& 
    \frac{\alpha}{\sqrt{\alpha^2+\nu^2}}\\
    \frac{\mu}{\sqrt{\alpha^2+\mu^2}}& 
    \frac{\nu}{\sqrt{\alpha^2+\nu^2}}
  \end{array}
\right)^{-1}
\end{equation}
For the simplicity we set 
\begin{equation}
P = 
\left(
  \begin{array}{cc}
    \frac{\alpha}{\sqrt{\alpha^2+\mu^2}}& 
    \frac{\alpha}{\sqrt{\alpha^2+\nu^2}}\\
    \frac{\mu}{\sqrt{\alpha^2+\mu^2}}& 
    \frac{\nu}{\sqrt{\alpha^2+\nu^2}}
  \end{array}
\right) \quad \in \quad O(2).
\end{equation}

\par \vspace{5mm} \noindent
In the text this is used as 
\begin{equation}
P_{n,\sigma} = 
\left(
 \begin{array}{cc}
  \frac{E_{\Delta,n,\sigma}}{\sqrt{E_{\Delta,n,\sigma}^2+\mu_{n,\sigma}^2}}& 
  \frac{E_{\Delta,n,\sigma}}{\sqrt{E_{\Delta,n,\sigma}^2+\nu_{n,\sigma}^2}}\\
  \frac{\mu_{n,\sigma}}{\sqrt{E_{\Delta,n,\sigma}^2+\mu_{n,\sigma}^2}}& 
  \frac{\nu_{n,\sigma}}{\sqrt{E_{\Delta,n,\sigma}^2+\nu_{n,\sigma}^2}}
 \end{array}
\right)
\quad \mbox{for}\quad \sigma=\pm 
\end{equation}
with 
\begin{equation}
\mu_{n,\sigma}
=\frac{1}{2}(\gamma+\sqrt{\gamma^{2}+4E_{\Delta,n,\sigma}^{2}}\ ),
\quad 
\nu_{n,\sigma}
=\frac{1}{2}(\gamma-\sqrt{\gamma^{2}+4E_{\Delta,n,\sigma}^{2}}\ )
\end{equation}
for $\gamma=\omega x^{2}$. Then 
\begin{equation}
{P_{n,+}}^{T}
\left(
  \begin{array}{cc}
    0 & 1  \\
    1 & 0
  \end{array}
\right)
P_{n,-}
= 
\left(
  \begin{array}{cc}
    a_{11} & a_{12} \\
    a_{21} & a_{22}
  \end{array}
\right), 
\end{equation}
where 
\begin{eqnarray}
a_{11}&=&
\frac{E_{\Delta,n,+}\mu_{n,-}+E_{\Delta,n,-}\mu_{n,+}}
{\sqrt{E_{\Delta,n,+}^2+\mu_{n,+}^2}\sqrt{E_{\Delta,n,-}^2+\mu_{n,-}^2}}, 
\quad 
a_{12}=
\frac{E_{\Delta,n,+}\nu_{n,-}+E_{\Delta,n,-}\mu_{n,+}}
{\sqrt{E_{\Delta,n,+}^2+\mu_{n,+}^2}\sqrt{E_{\Delta,n,-}^2+\nu_{n,-}^2}},  
\nonumber \\
a_{21}&=&
\frac{E_{\Delta,n,+}\mu_{n,-}+E_{\Delta,n,-}\nu_{n,+}}
{\sqrt{E_{\Delta,n,+}^2+\nu_{n,+}^2}\sqrt{E_{\Delta,n,-}^2+\mu_{n,-}^2}}, 
\quad 
a_{22}=
\frac{E_{\Delta,n,+}\nu_{n,-}+E_{\Delta,n,-}\nu_{n,+}}
{\sqrt{E_{\Delta,n,+}^2+\nu_{n,+}^2}\sqrt{E_{\Delta,n,-}^2+\nu_{n,-}^2}}.  
\nonumber 
\end{eqnarray}

\newpage

%%%%%%%%%%%%%
%References%
%%%%%%%%%%%%%

\end{document}